\documentclass[a4paper,fleqn,usenatbib,useAMS]{mnras}
\usepackage{mathptmx}

\usepackage[T1]{fontenc}
\usepackage{ae,aecompl}
\usepackage{aecompl}

\usepackage{graphicx}	
\usepackage{amsmath}	
\usepackage{amssymb}	

\usepackage{times}
\usepackage{epstopdf}
\usepackage{graphicx}
\usepackage{txfonts}
\usepackage{natbib}
\usepackage{array}
\title[Beware the effect of edge-on discs]{SHARP - II. Mass structure in strong lenses is not necessarily dark matter substructure: A flux ratio anomaly from an edge-on disc }
\author[Hsueh et al.]{J.-W. Hsueh,$^{1}$\thanks{E-mail: jwhsueh@ucdavis.edu} C. D. Fassnacht,$^{1}$ S. Vegetti,$^{2}$  J. P. McKean,$^{3,4}$ C. Spingola,$^{4}$ M. W. Auger,$^{5}$
\newauthor  L. V. E. Koopmans,$^{4}$ and D. J. Lagattuta$^{6}$\\
$^{1}$Department of Physics, University of California, Davis, 1 Shields Ave. Davis, CA 95616, USA\\
$^{2}$Max Planck Institute for Astrophysics, Karl-Schwarzschild-Strasse 1, D-85740 Garching, Germany\\
$^{3}$Netherlands Institute for Radio Astronomy (ASTRON), P.O. Box 2, 7990 AA Dwingeloo, The Netherlands\\
$^{4}$Kapteyn Astronomical Institute, University of Groningen, P.O. Box 800, 9700 AV Groningen, The Netherlands\\
$^{5}$Institute of Astronomy, University of Cambridge, Madingley Road, Cambridge CB3 0HA, UK\\
$^{6}$University of Lyon, CRAL, Observatoire de Lyon, 92 Rue Pasteur, 69007 Lyon, France}
\begin{document}


\pagerange{\pageref{firstpage}--\pageref{lastpage}} \pubyear{2015}

\maketitle

\label{firstpage}

\begin{abstract}
Gravitational lens flux-ratio anomalies provide a powerful technique for measuring dark matter substructure in distant galaxies. However, before using these flux-ratio anomalies to test galaxy formation models, it is imperative to ascertain that the given anomalies are indeed due to the presence of dark matter substructure and not due to some other component of the lensing galaxy halo or to propagation effects. Here we present the case of CLASS~B1555+375, which has a strong radio-wavelength flux-ratio anomaly.  Our high-resolution near-infrared Keck~II adaptive optics imaging and archival {\it Hubble Space Telescope} data reveal the lensing galaxy in this system to have  a clear edge-on disc component that crosses directly over the pair of images that exhibit the flux-ratio anomaly. We find that simple models that include the disc can reproduce the cm-wavelength flux-ratio anomaly without requiring additional dark matter substructure. Although further studies are required, our results suggest the assumption that all flux-ratio anomalies are due to a population of dark matter sub-haloes may be incorrect, and analyses that do not account for the full complexity of the lens macro-model may overestimate the substructure mass fraction in massive lensing galaxies.
\end{abstract}

\begin{keywords}
gravitational lensing -- quasars: individual CLASS B1555+375 -- galaxies: structure 
\end{keywords}

\section{Introduction}

A common feature of dark-matter only simulations of structure formation is the presence of thousands of subhaloes that are associated with larger mass haloes \citep[e.g.][]{Springel08}. However, observations of the Local Group show many fewer satellite galaxies than are predicted by the simulations, even when taking into account the latest discoveries of new dwarf galaxies \citep{DES15,Kop15} and completeness corrections arising from the limited sky coverage imposed by the Galactic Plane.  
This is the well-known {\it missing satellite problem} \citep{Klypin1999,Moore1999,S07}. In order to understand this discrepancy, it is imperative to build up a large sample of satellites/subhaloes in galaxies outside the Local Group. However, this process is challenging due to the 
faintness of the satellite galaxies.  In fact, lower-mass satellites may not be able to retain the gas needed for ongoing star formation \citep[e.g.][]{P11}, rendering them effectively dark.  

\begin{figure*}
\includegraphics[width=0.95\textwidth]{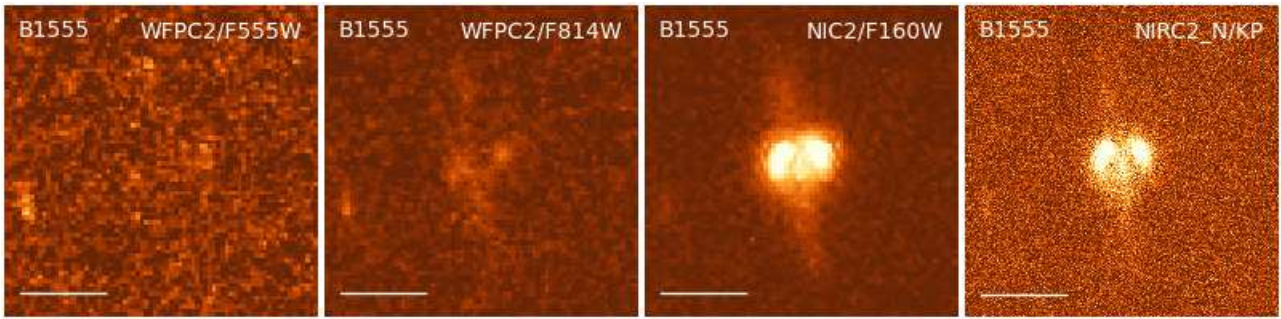}
\caption{High-resolution multi-band imaging of B1555+375 with {\it HST}/WFPC2 at F555W (left) and F814W (middle-left), with {\it HST}/NICMOS/NIC2 at F160W (middle-right) and with Keck adaptive optics at $K^\prime$-band (right). All of the panels are 3.7 arcsec on a side and the scale-bar represents 1 arcsec.}
\label{fig:multiband}
\end{figure*}

It was first suggested by \citet{Mao1998} that the flux-ratio anomalies observed in the multiple radio-loud images of lensed quasars could be interpreted as a sign of the presence of substructure in lens galaxies. Indeed, small perturbations in the gravitational potential are sufficient to cause flux-ratio anomalies in gravitationally lensed systems \citep{metcalf01,Dalal2002,Bradac02}, particularly for those systems with a close pair of merging images \citep{KD04}. At radio wavelengths, the lensed images are neither sensitive to micro-lensing from stars (\citealt{K03}; although see \citealt{koopmans00}), nor to dust extinction. Also, any variation in the properties of the radio images that are due to propagation effects (e.g. interstellar scattering or free-free absorption) have a well understood frequency dependence that can be identified and accounted for \citep[e.g.][]{biggs03,M07,winn04}. This makes the flux-ratio anomalies of radio-loud lensed images a promising tool to detect and measure dark matter substructures in a large sample of distant galaxies \citep{Dalal2002}. However, a recent comparison of the observational data (eight gravitational lens systems with four or more images) with the predictions from a set of cold dark matter simulations that are representative of the mass and redshift of the lens galaxies suggests that there is only a 1--4 per cent probability that the full observed flux-ratio distribution is produced only by substructure \citep{Xu15}.

The \emph{gravitational imaging technique}, which alternatively utilizes the surface brightness distribution of extended arcs and Einstein rings to detect low-mass substructures within lensing haloes, was introduced by \citet{K05} and \citet{V09}. This method has given a measurement of the substructure mass fraction for a sample of 11 lenses that is currently in agreement with the theoretical expectations from cold matter matter simulations \citep{V14a,V12}. The reason for the conflicting results between the flux-ratio anomaly and gravitational imaging techniques is currently not clear.  While this could be the result of small sample statistics, it could also be an indication that the flux-ratio anomalies have a different origin than clumpy substructure \citep[see][for a discussion]{Xu15}.

 
In this letter, we explore the idea that not all flux-ratio anomalies are due to the presence of clumpy substructure, but that some originate from more complex lens galaxy mass distributions than have initially been considered. To this end, we use new infrared and radio interferometric imaging data from the Strong lensing at High Angular Resolution Programme \citep[SHARP;][Chen et al. 2016 submitted]{mckean07,lagattuta10,SHARP12, V12}. The main goal of SHARP is to investigate galaxy structure and other astrophysical topics through the re-observation of known gravitational lens systems at high angular resolution, making use of observations with (1) adaptive optics (AO) on the W. M. Keck telescope, (2) the {\it Hubble Space Telescope (HST)}, and (3) large radio and mm interferometric arrays. 

Here we focus on CLASS B1555+375 \citep{Marlow99}, which was discovered as part of the Cosmic Lens All-Sky Survey \citep{CLASS1,CLASS2}. B1555+375 has four lensed images with a maximum separation of 426~mas in a fold configuration.  The two merging lensed images show a strong and persistent flux-ratio anomaly at radio wavelengths, and previous radio monitoring observations ruled out microlensing effects as the cause of the anomaly \citep{K03}. As B1555+375 was included in the observational and theoretical analyses of both \citet{Dalal2002} and \citet{Xu15}, understanding its mass distribution is of key importance to assess any systematics with the inferred level of substructure in the lensing halo with the flux-ratio anomaly method. The SHARP imaging data are introduced in Section 2 and we present a revised lens model using these data in Section 3. The results of this analysis and their implications are discussed in Section 4. Note that the redshifts of both the lens galaxy and the lensed source are unknown for B1555+375, so we assume redshifts of $z_l = 1$ and $z_s = 1.5$, respectively, due to their red colours \citep{Marlow99}. This assumption has no effect on the derived magnifications or flux ratios of our analysis.

\section{Observations \& Data reduction}

Our dataset consists of optical and infrared imaging taken with the W. M. Keck-II Telescope and HST, and high-resolution radio imaging taken with the Very Long Baseline Array (VLBA), which are discussed below and are summarized in Table~\ref{tab:obs}.

\subsection{Keck adaptive optics and {\it HST} imaging}

B1555+375 was observed as part of SHARP using the NIRC2 camera on the W. M. Keck-II Telescope on 2012 May 16 (PI: Fassnacht).  The AO system was used, with the corrections derived from the laser guide star and a $R=14.4$-mag tip-tilt star located $\sim45$~arcsec from the lens.  The narrow camera mode was used, giving a field of view of $\sim10$~arcsec on a side and a pixel scale of 10~mas. Six dithered 300~s exposures were obtained in the $K^{\prime}$ filter.  The data were reduced with the standard SHARP pipeline, which is a python-based package that is a refinement of the process described by \citet{Auger_EELS1}.  A cutout of the final reduced image is shown in Fig.~\ref{fig:multiband} and also in Fig.~\ref{fig:3panel}a, where the contours from 5-GHz Multi-Element Radio Link Interferometer Network (MERLIN) imaging by \citet{Marlow99} and 1.66-GHz VLBA imaging (see below) are overlaid.

B1555+375 was also observed with the {\it HST} in three broad bands.  The optical data were obtained with the Wide-Field Planetary Camera 2 (WFPC2) in the F555W and F814W bands (GO-8804; PI: Falco), while the Near Infrared Multi-Object Spectrograph (NICMOS) was used to observe the system in the F160W band (GO-9744; PI: Kochanek).  The NICMOS observations were obtained with the NIC2 camera.  We reduced all of the archival \textit{HST} data with the standard {\sc multidrizzle} pipeline, producing final drizzled images with pixel-sizes of 50~mas. The reduced images are also shown in Fig.~\ref{fig:multiband}. The lens and the lensed images are not detected at high significance in the optical bands, but are clearly seen in the NICMOS F160W image.

There are several notable features in the high-resolution AO and {\it HST} data. These include the nearly complete lack of emission associated with lensed images B and D, 
(Fig.~\ref{fig:3panel}b), and the faint, but clearly visible emission from the lensing galaxy that appears to be an edge-on disc, with a position angle of $\sim 10$~deg and an axis ratio $q \sim 0.2$.  We note that the position angle of the lensing galaxy is such that the disc emission lies close to the lensed images B and D. Therefore, the lack of emission from these rest-frame optical images is likely due to dust extinction and additional demagnification due to lensing. 

\subsection{Very Long Baseline Array imaging}

B1555+375 was observed with the 10 telescopes of the VLBA at a central observing frequency of 1.66 GHz on 2000 March 21 (BN0009; PI: Norbury). The data were recorded at 128 Mbits~s$^{-1}$ and then correlated to produce two spectral windows with 8 MHz bandwidth, 16 channels and both circular polarizations (RR and LL). The observations were phase referenced using J1544+398 every 3 min over the total observing time of 3 h. The data were reduced in the standard manner using the {\sc vlbautlis} analysis pipeline that is part of the Astronomical Image Processing Software (AIPS). Imaging was done with the {\sc clean} algorithm in AIPS using natural weighting of the visibilities, and restored using an elliptical beam of size $9.7 \times 6.9$~mas at a position angle of $-7.6^\circ$ east of north. The final map (also shown in Fig.~\ref{fig:3panel}a) has an rms of 78~$\mu$Jy~beam$^{-1}$.

We find that lensed images A and B have been resolved into a gravitational arc that is $\sim100$~mas in length. Image C is still compact on mas-scales, whereas image D is only marginally detected ($\sim6~\sigma$). 
The VLBA imaging shows that image B has a smaller angular size than image A, which is consistent with a perturbation in the mass model being responsible for the flux-ratio anomaly (recall that gravitational lensing conserves surface brightness), as opposed to free-free absorption or interstellar scattering.

\begin{table}
\centering
\caption{Summary of the B1555+375 observations.}
\begin{tabular}{llccc}
\hline
Telescope		& Camera			&  Band 		& Date				&$t_{exp}$ (s) \\
\hline
VLBA			&					& 1.66 GHz	& 	2000 Mar 21	& 10800\\
\textit{HST}	& WFPC2    		& F555W		&	2000 Oct 09	& 5200\\
\textit{HST}	& WFPC2    		& F814W		&	2000 Oct 09 	& 5200\\
\textit{HST}	& NICMOS/NIC2	& F160W		&	2003 Nov 02	& 5376\\
Keck-II			& NIRC2 AO		& $K^\prime$	& 	2012 May 16	& 1800\\
\hline
\end{tabular}
\label{tab:obs}
\end{table}

\section{Lens modelling}

The flux-ratio anomaly and relative sizes of the images at radio-wavelengths imply that there must be some form of perturbation to the gravitational lens mass model, which has thus far been attributed to dark matter substructure within the lensing galaxy or along the line of sight to the lensed images \citep{Dalal2002,Xu12,Xu15}. However, our high-resolution near infrared imaging suggests the intriguing possibility that the mass perturbation could be due to an edge-on disc component within the lens. To test this possibility, we model the system with a disc component in order to see if a plausible mass model can explain the flux-ratio anomaly without the need for additional substructure.

We use the lens modelling code {\sc gravlens} \citep{Kee01} to model the compact radio components. The inputs to the model are the observed image positions measured from the MERLIN radio observations of \citet{Marlow99} as all of the images are significantly detected in these data. The flux-ratio measurements are taken from \citet{K03}, who used half a year of MERLIN monitoring to obtain flux-ratio curves. In total, there are 11 constraints to the mass model provided by the observational data. As a first trial of modelling, we re-create the singular isothermal ellipsoid without external shear model (SIE; 7 free parameters) of \citet{Marlow99} to check the performance of a simple lensing potential. Consistent with previous studies \citep{Marlow99,Xu15}, a single elliptical potential model cannot reproduce the observed flux ratios in B1555+375 (see Table~\ref{tab:results}). 

The next step is to test if a physically plausible edge-on disc can cause a perturbation in the strong lensing mass model that produces the observed flux ratio anomaly of  merging images A and B. An exponential disc profile best describes the disc component in spiral galaxies because it matches the light distribution \citep{Kee98}. We thus choose a SIE plus an exponential disc model (SIE+ExpDisc) in our study of B1555+375. The free parameters of this model are the SIE Einstein radius ($b$), centroid position, ellipticity ($e = 1 - q$), position angle ($\theta$), the exponential disc intrinsic central density ($\kappa_0$) and scale length ($R_s$). A detailed definition of these mass profiles is described by \citet{Kee01}. The source position adds two additional model parameters. We have not included a third component such as an external shear or additional NFW halo term because the number of constraints provided by the compact images detected by MERLIN is not sufficient to constrain such a model.

More sophisticated mass models require additional
constraints, such as those from the lensed extended emission in either
the NIR or radio imaging.  While we have successfully used the former
approach in the past \citep[e.g.][]{V12}, for B1555+375 the extinction
in the AO imaging is so severe that components B and D are not
visible, and the arc that connects A and B appears to be truncated.
The low signal-to-noise ratio in the ring away from the lensed AGN and
the strong extinction in the lens galaxy prevent us from using this
gravitational imaging approach.  Significantly deeper
observations are needed at a higher bandwidth with very long baseline
interferometric arrays, which can test more complex models in the
future by imaging the extended gravitational arc and more
significantly detecting image D. 

We use {\sc gravlens} to obtain the best-fit SIE+ExpDisc lens model. The best-fit parameters from \citet{Marlow99} are chosen as the starting point of the SIE profile. We have used the NICMOS F160W imaging, in which the disc component has the highest signal-to-noise ratio amongst our data, to fix the disc centroid position relative to image A to $(-0.162, -0.206)$, its ellipticity to $e=0.83$, and its position angle to $\theta=8^\circ$. The total number of free parameters is 9. After obtaining the best-fit lens model, we run a Markov-Chain Monte-Carlo (MCMC) analysis to obtain the uncertainties.

\begin{figure*}
\includegraphics[width=0.9\textwidth]{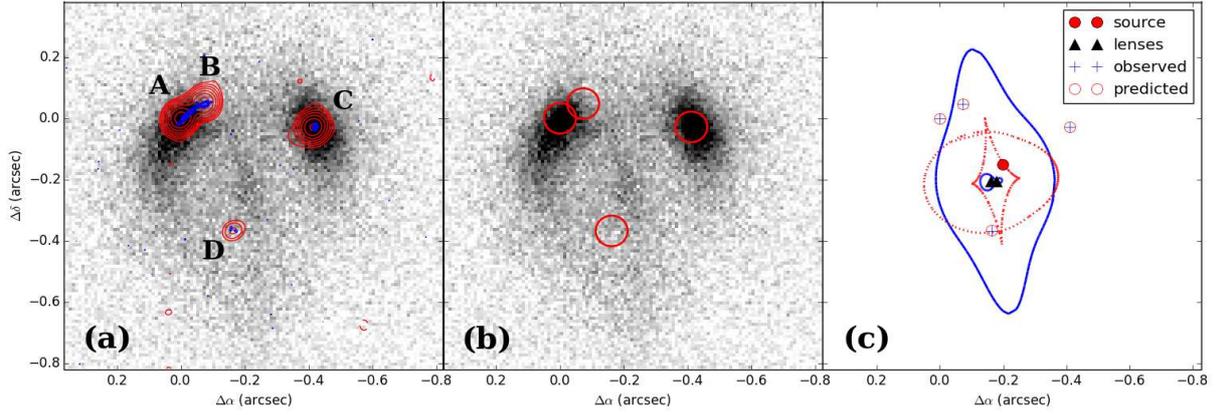}
\caption{
{\bf (a)} $K^\prime$-band AO imaging of B1555+375 with contours from the MERLIN observations (red) of \citet{Marlow99} and the VLBA observations presented here (blue) overlaid.
{\bf (b)} $K^\prime$-band AO image with the locations of the four radio components
marked with circles.  
{\bf (c)} Observed radio (blue pluses) and model-predicted (red open
circles) image positions of B1555+375. The lens-plane critical curves
are shown with the blue solid line and the source-plane caustics are
shown with the red dotted line. The position of the source is at
($-0.1953$, $-0.1494$), which is marked by the red filled circle. The
centroid positions of the two mass components of the lens (a SIE and
an exponential disc) are marked by the black filled triangles.
}
\label{fig:3panel}
\end{figure*}

The best-fit model parameters and 68 per cent confidence level (CL) errors are listed in Table~\ref{tab:model} and the lens model is illustrated in Fig.~\ref{fig:3panel}c. Table \ref{tab:results} also shows the comparison between the observed radio data and the model-predicted image positions and flux ratios. We find that by adding an elongated exponential disc to a smooth SIE model there is significant improvement in the predicted flux ratios when compared to the results for the SIE-only model. 
The positions are all in agreement, and the flux ratios are consistent at the $2\sigma$ level, with excellent agreement now obtained for the two merging images A and B. The remaining differences in the fluxes may be due to the real lensed source being slightly extended rather than the point source that we used in our model. The model requires that the ExpDisc component makes up about 15 per cent of the total projected mass within the Einstein radius (1.7~kpc for an assumed lens redshift of $z_l = 1$ but the mass fraction is independent of the lens and source redshifts since this ratio is independent of distance).  In this central region of an edge-on spiral galaxy, the baryonic mass can be a significant component of the total projected mass. Investigations of edge-on spiral lenses, for example from the SWELLS survey, show that the disc mass fraction within the Einstein radius ranges from 5 to 30 per cent\citep{swells1,swells3,swells6}.

%

\begin{table*}
\centering
\caption{A summary of the observational constraints and the predicted lens-model parameters. The lensed image positions in the radio \citep{Marlow99} (in arcsec) and average flux ratios \citep{K03} are with respect to image A. Uncertainties in the position measurements for images A, B and C are $0.001$~arcsec and for image D are $0.006$~arcsec.}
\begin{tabular}{cccccccc}
\hline
Image	&\multicolumn{3}{c}{Observed} 	 	& \multicolumn{4}{c}{Model-Predicted}\\
		&\multicolumn{3}{c}{MERLIN 5 GHz}		& & & {SIE+ExpDisc} & SIE-only\\
		&East &North & Flux ratio &East 	&North & Flux ratio &Flux ratio\\ 
\hline
A  &$0$    		&$0$			&  $1$ 				&$+0.000$  &$+0.000$	& $1$ 		& $1$\\  
B  &$-0.0726$	&$+0.0480$	& $0.620 \pm 0.039$ 	&$-0.0715$ &$+0.0487$		& $0.606$ 	& $0.971$ \\  
C  &$-0.4117$ 	&$-0.0280$	& $0.507\pm 0.030$	&$-0.4115$ &$-0.0278$		& $0.445$ 	& $0.312$\\  
D  &$-0.1619$	&$-0.3680$	& $0.086 \pm 0.024$ 	&$-0.1693$ &$-0.3704$		& $0.123$ 	& $0.106$\\  
\hline
\end{tabular}
\label{tab:results}
\end{table*}

\begin{table}
\centering
\caption{Best-fit model parameters and uncertainties for our two-component SIE + exponential disc lens model (our single SIE model is effectively equivalent to the model in \citet{Marlow99} and so is not shown here). The positions are measured in arcsec, relative to the MERLIN radio position of image A. The ellipticity is defined as $e=1-q$ and the position angles are measured in degrees east of north. The Einstein radius $b$ and disc scale radius ($R_s$) are measured in arcsec.}
\begin{tabular}{ccc}
\hline 
Parameter    & SIE Component & ExpDisc Component  \\
\hline
$\Delta$RA	& $-0.177 \pm 0.003$		& $-0.162$ \textit{(fixed)}\\
$\Delta$Dec	& $-0.205 \pm 0.004$		& $-0.206$ \textit{(fixed)} \\
$b$ 			& $0.178 \pm 0.004$  		& ...  \\
$\kappa_0$ & ... &$0.273 \pm 0.003$\\
$e$	  		& $0.25 \pm 0.05$			& $0.83$ \textit{(fixed)} \\
$\theta$ 		& $4.9 \pm 1.3$			& $8.0$ \textit{(fixed)}	 \\
$R_s$			& ...  						& $0.236 \pm 0.003$	 \\
\hline
\label{tab:model}
\end{tabular}
\end{table}

\section{Discussion \& Conclusions}

Although flux-ratio anomalies provide a powerful tool to investigate low-mass substructure in lensing haloes out to cosmological redshifts, it is clear that some care is needed in interpreting the results. It is now well established that observations of radio-loud gravitational lenses may provide the most robust observational constraints because they are less affected by dust extinction and microlensing, and monitoring can determine the best estimate of the intrinsic flux ratios. This has led to an oft-studied sample of eight radio-loud gravitational lenses with well-defined flux-ratio measurements \citep{Dalal2002,KD04,Xu15}. In general, lens models that are based only on the locations of lensed radio-loud images have large degeneracies due to the limited number of constraints that are provided by 50 to 200~mas imaging of the unresolved sources \citep[e.g.][]{Ka91}. Furthermore, observed flux-ratio anomalies may not be due exclusively to a population of low-mass "dark" substructures, as already seen in the MG J0414+0534 and MG J2016+112 systems, where there is evidence for luminous companion dwarf galaxies that can account for the observed flux ratios, and even observed astrometric anomalies (\citealt{ros00,chen07,more09}; see also \citealt{mckean07,jackson10}). 

Here, we find evidence for an edge-on disc in the flux-ratio anomaly
gravitational lens B1555+375 that can account for the required
perturbation in the mass model.  Although a constraint on the disc mass from photometry is currently not available due to the low signal-to-noise ratio in the AO imaging and the lack of redshift information, our model indicates that the flux-ratio anomaly in B1555+375 has a strong link to the presence of an edge-on disc.  Such
high-order deviations from a smooth mass distribution have already
been suggested previously \citep{evans03, congdon05}, but this is the
first time that an observed disc has been explicitly modelled to
account for a flux-ratio anomaly. However, it is important to realise
that by including observationally-motivated modifications to a
`smooth' mass distribution, such as luminous dwarf companions or
edge-on discs, in the mass models, we are still limited in our
interpretation if we can not determine how much of the anomaly are
due to these components. For our current analysis of B1555+375, we
have shown that the entire anomaly could be caused by the edge-on
disc. 
That is not to imply this is a unique model, but rather that the
disc component is a plausible explanation of the anomaly. 

In future it may be possible to independently constrain the mass
of the disc through kinematic measurements (most likely from
mm-wavelength resolved emission line observations), which would test
whether the disc actually makes a sufficient contribution to the mass
model \citep[e.g. as in the case of B1933+503;][]{suyu12}. Also, in
the blue contours of Fig.~\ref{fig:3panel}a, B1555+375 shows evidence
for an extended $\sim$ 100 mas gravitational arc at radio wavelengths
that crosses the edge-on disc. Further high resolution imaging with
very long baseline interferometry will provide the observational
constraints required to fully test whether the disc is sufficient to
cause the anomaly.

Our analysis of B1555+375 demonstrates that the choice of the macromodel used to describe the `smooth' lens galaxy mass distribution, combined with the limited information provided by the position of compact lensed images, can lead to a misinterpretation of the origin of flux-ratio anomalies in some systems. Special care may be needed for small-separation lens systems, in which the lensing galaxy is likely to be a spiral \citep[e.g.,][]{Turner84, F92}.  This is because a disc with a high inclination angle can significantly affect the lensing cross-section for such systems \citep{Maller97,Kee98,Moller03}. The current sample of lensed quasars with radio-wavelength flux ratio measurements is small.  Therefore, having even a few systems with edge-on discs may be sufficient to introduce a bias into the substructure analysis. In fact, our near-infrared AO imaging of
other lenses in the SHARP sample shows that B1555+375 is not the only
system that has a significant disc component that crosses the merging lensed images (Fassnacht et al., in prep.).  
These gravitational lenses will be investigated in future
work, but the images alone suggest that other systems may have a similar
link between flux-ratio anomalies and disc components. This potential bias introduced by edge-on discs may be avoided by post-selection as  sample sizes grow in the future. Therefore, it is critical to obtain high-resolution imaging lenses to detect additional mass features such as edge-on discs or even massive luminous satellites.
 This result is timely, as upcoming large-scale surveys (e.g., the Dark Energy Survey, Kilo-Degree Survey, \textit{Euclid}, and the Large Synoptic Survey Telescope) will deliver orders of magnitude more gravitationally lensed quasars for which flux ratios can
  be measured using narrow lines \citep[e.g.][]{N14}.

\section*{Acknowledgments}
We thank the referee for comments that improved the paper.  We are grateful to Dandan Xu for useful discussions. CDF and DJL acknowledge support from NSF-AST-0909119.  LVEK is supported in part through an NWO-VICI career grant (project number 639.043.308). The NRAO is a facility of the NSF operated under cooperative agreement by Associated Universities, Inc. Based on observations made with the NASA/ESA Hubble Space Telescope, obtained from the data archive at the Space Telescope Science Institute. STScI is operated by the Association of Universities for Research in Astronomy, Inc. under NASA contract NAS 5-26555. Some of the data presented herein were obtained at the W. M. Keck Observatory, which is operated as a scientific partnership among the California Institute of Technology, the University of California and the National Aeronautics and Space Administration. The Observatory was made possible by the generous financial support of the W. M. Keck Foundation. The authors wish to recognize and acknowledge the very significant cultural role and reverence that the summit of Mauna Kea has always had within the indigenous Hawaiian community.  We are most fortunate to have the opportunity to conduct observations from this mountain.

\bibliographystyle{mnras}
\bibliography{reference}

\bsp
\label{lastpage}

\end{document}